# Magnetic Shape Memory Polymers with Integrated Multifunctional Shape Manipulations

Qiji Ze[a§], Xiao Kuang[b§], Shuai Wu[a§], Janet Wong[b], S. Macrae Montgomery[b], Rundong Zhang[a], Joshua M. Kovitz[c], Fengyuan Yang[d], H. Jerry Qi[b]*, Ruike Zhao[a]*

[a]Department of Mechanical and Aerospace Engineering, The Ohio State University, Columbus, OH, 43210, USA
[b]The George W. Woodruff School of Mechanical Engineering, Georgia Institute of Technology, Atlanta, GA 30332, USA
[c]Georgia Tech Research Institute, Atlanta, GA 30332, USA
[d]Department of Physics, The Ohio State University, Columbus, OH, 43210, USA

[§] These authors made equal contribution to this work.
*Corresponding authors. Email: qih@me.gatech.edu; zhao.2885@osu.edu

**Shape-programmable soft materials that exhibit integrated multifunctional shape manipulations, including reprogrammable, untethered, fast, and reversible shape transformation and locking, are highly desirable for a plethora of applications, including soft robotics, morphing structures, and biomedical devices. Despite recent progress, it remains challenging to achieve multiple shape manipulations in one material system. Here, we report a novel magnetic shape memory polymer composite to achieve this. The composite consists of two types of magnetic particles in an amorphous shape memory polymer matrix. The matrix softens via magnetic inductive heating of low-coercivity particles, and high-remanence particles with reprogrammable magnetization profiles drive the rapid and reversible shape change under actuation magnetic fields. Once cooled, the actuated shape can be locked. Additionally, varying the particle loadings for heating enables sequential actuation. The integrated multifunctional shape**

**manipulations are further exploited for applications including soft magnetic grippers with large grabbing force, sequential logic for computing, and reconfigurable antennas.**

**Introduction**

Shape programmable soft materials that exhibit integrated multifunctional shape manipulations, including reprogrammable, untethered, fast, and reversible shape transformation and locking, in response to external stimuli, such as heat, light, or magnetic field[1-5], are highly desirable for a plethora of applications, including soft robotics[6], actuators[7-9], deployable devices[10,11], and biomedical devices[6,12-15]. A wide range of novel materials have been developed in the past, including liquid crystals elastomers[16,17], hydrogels[18], magnetic soft materials[6,19], and shape memory polymers (SMPs)[1,20-22]. Magnetic soft materials composed of magnetic particles in a soft polymer matrix have drawn great interest recently due to their untethered control for shape change[23,24], motion[6,7,25], and tunable mechanical properties[26]. Among them, hard-magnetic soft materials utilize high-remanence, high-coercivity magnetic particles, such as neodymium-iron-boron (NdFeB), to achieve complex programmable shape changes[6,19,27-29]. Under an applied magnetic field, these particles with programmed domains exert micro-torques, leading to a large macroscopic shape change. However, maintaining the actuated shape needs a constantly applied magnetic field, which is energy inefficient. In many practical applications, such as soft robotic grippers[30,31] and morphing antennas[32,33], it is highly desirable that the actuated shape can be locked so that the material can fulfill certain functions without the constant presence of an external field.

SMPs can be programmed and fixed into a temporary shape and then recover the original shape under external stimuli, such as heat or light [34,35]. Typically, a thermally triggered SMP uses a transition temperature ($T_{tran}$), such as glass transition temperature ($T_g$), for the shape memory effect. In a shape memory cycle, an SMP is programmed to a temporary shape by an external force at a temperature above $T_{tran}$ followed by cooling and unloading. The SMP recovers its original shape at temperatures above $T_{tran}$ by direct heating or inductive heating[36,37].

In this article, motivated by the advantages of hard-magnetic soft materials and SMPs, we report a novel magnetic shape memory polymer (M-SMP) with integrated reprogrammable, untethered, fast, and reversible actuation and shape locking. The M-SMP is composed of two types of magnetic particles ($Fe_3O_4$ and NdFeB) in an amorphous SMP matrix. The $Fe_3O_4$ particles enable inductive heating under a high frequency alternating current (AC) magnetic field and thus are employed for shape locking and unlocking. The NdFeB particles are magnetized and remagnetized with predetermined magnetization profiles for programmable actuation. We demonstrate that the integrated multifunctional shape manipulations offered by M-SMPs can be exploited for a wide range of novel applications, including soft grippers for heavy loads, sequential logic circuits for digital computing, and reconfigurable morphing antennas.

**Results**

**Design and characterization**

To demonstrate the concept, we fabricate an acrylate-based amorphous SMP with embedded NdFeB microparticles and $Fe_3O_4$ microparticles (**Methods**, **Figs. S1**-**S3, Table S1 & S2**). Before use, the M-SMP is magnetized to have a desired magnetic profile under an impulse magnetic field (~1.5 T). **Fig. 1a** shows the working mechanism by using an M-SMP cantilever with a magnetization polarity along its longitudinal direction. At room temperature, the cantilever is stiff and cannot deform under an actuation magnetic field ($B_a$). When an AC magnetic field ($B_h$) is applied, the inductive heating of the $Fe_3O_4$ particles heats the M-SMP above its $T_g$, and the modulus of the M-SMP drops significantly. Then, a small $B_a$ can bend the cantilever. By alternating $B_a$ between up (+) and down (-) directions at this moment, fast transforming between upward and downward bending can be easily achieved. Upon removal of $B_h$, the bending shape can be locked without further applying $B_a$ once the temperature of the M-SMP drops below its $T_g$. Moreover, the magnetization profile of the M-SMP can be reprogrammed for different shape transformation by remagnetization. For example, remagnetizing the beam when it is mechanically locked in a folding shape will change the actuation shape to folding under the same $B_a$ (bottom row of **Fig. 1a**).

Neat SMP and M-SMP samples are prepared to characterize their thermomechanical properties. **Fig. 1b** shows the thermomechanical properties of the neat SMP and the M-SMP P15-15, where the two numbers represent the volume fractions of $Fe_3O_4$ and NdFeB particles, respectively. The storage modulus of P15-15 decreases from 4.6 GPa to 3.0 MPa when the temperature $T$ increases from 20°C to 100°C. $T_g$, measured as the temperature at the peak of the tanδ curve, is ~56°C for the neat SMP, and ~58°C for P15-15 (**Fig. S4**). The Young's modulus of the M-SMP at high-temperature increases linearly with the increasing particle loading (**Fig. 1c**). **Fig. 1d** shows the strain, stress, and temperature as functions of time during the shape memory test of P15-15. When P15-15 is programmed at 85°C, it has the shape fixity and shape recovery ratios of 87.8% and 87.2%, respectively (**Fig. S5**).

The $Fe_3O_4$ particles, due to their low coercivity, can be easily magnetized and demagnetized under a small high frequency AC magnetic field, leading to a magnetic hysteresis loss for inductive heating. In contrast, the NdFeB particles, due to their high coercivity, can retain high remnant magnetization for magnetic actuation (**Fig. S6 & S7**). Note that the NdFeB particles start to be demagnetized when the temperature is above ~150ºC (**Fig. S8**). Therefore, the temperature for shape unlocking and actuation should be limited to below 150ºC.

**Fast transforming and shape locking**

Here, we experimentally demonstrate the remote fast transforming and shape locking of the M-SMP, which can be used as a soft robotic gripper. The experimental setup for M-SMP heating and actuation consists of two types of coils (**Fig. 2a**): a pair of electromagnetic coils generate $B_a$ for actuation; a solenoid provides $B_h$ for inductive heating. An M-SMP (P15-15) cantilever is fabricated with magnetization along its longitudinal direction in such a way that the beam will tend to bend under a vertical magnetic field (**Fig. 2b**). To heat and actuate the beam, we use $B_h$ = 40 mT at 60 kHz and $B_a$ = 30 mT, respectively. The magnetic field profiles for $B_a$ and $B_h$, as well as the measured cantilever displacement versus time, are shown in **Fig. 2c**. The application of $B_h$ gradually increases the temperature and the deflection of the M-SMP (**Supplementary Video S1**). Here, we alternate $B_a$ at 0.25 Hz to show the reversible fast transforming. Upon removal of $B_h$ at 30 s, the temperature drops by air cooling and the modulus of the M-SMP increases dramatically (**Fig. 1b**). The bending shape can then be locked without further application of $B_a$. **Fig. 2d** shows the M-SMP cantilever carrying a weight (23 g) that is 64 times heavier than its own weight (0.36 g).

Soft robotic grippers are intensively researched due to their capability of adapting their morphology to grab objects. However, the low-stiffness nature of soft materials significantly limits the actuation force, making most soft robotic grippers incapable of grabbing heavy objects. Taking M-SMPs' advantage of shape locking, we demonstrate a soft robotic gripper that grabs an object much heavier than its own weight. **Fig. 2e** shows the design and magnetization directions of a four-arm gripper (**Fig. S9**). By applying $B_h$ and a positive $B_a$ (upward), the gripper softens and opens up for grabbing. Upon switching $B_a$ to negative, the gripper conforms to the lead ball. At this moment, the ball slips if the gripper is lifted (**Fig. 2f**). However, the gripper can be locked into the actuated shape and provide a large grabbing force when we remove $B_h$ and cool down the material. As demonstrated in **Fig. 2g**, the stiffened gripper can effectively lift the lead ball without any external stimulation (**Supplementary Video S2**). The weight of the lead ball is 23 g, which is 49 times heavier than the gripper (0.47 g).

**Sequential actuation**

The sequential shape transformation of an object in a predefined sequence can enable a material or system to fulfill multiple functions[25,38]. Here, we show that the sequential actuation of an M-SMP system can be achieved by designing and actuating material regions with different $Fe_3O_4$ loadings for different resultant heating temperatures and stiffnesses under the same applied $B_h$. We prepare three M-SMPs with the same dimension containing the same amount of NdFeB (15 vol%) but different amounts of $Fe_3O_4$ (5 vol%, 15 vol%, and 25 vol%, named as P5-15, P15-15, and P25-15, respectively). **Fig. 3a** shows the mechanical and heating characterizations of the three M-SMPs under the same $B_h$ (**Methods, Fig. S10**). To reach the temperature (around 50°C) at which the M-SMPs become reasonably soft to deform under $B_a$, it takes 5 s, 11 s, and 35 s for P25-15, P15-15, and P5-15, respectively.

Based on the mechanism of sequential actuation, we design a flower-like structure made of M-SMP petals using P5-15 and P25-15 to demonstrate the programmable sequential motion (**Fig. 3b**). The P5-15 petals are designed to be longer than the P25-15 ones, and the magnetization is along the outward radial direction for all petals (**Fig. S11**). **Fig. 3c** shows $B_h$ (red) and $B_a$ (black) profiles as functions of time. The deflections of P5-15 and P25-15 petals, defined as the vertical displacements of the endpoints, are plotted as black and blue curves in **Fig. 3c**, with the sequential shape change illustrated in **Fig. 3d**. Upon the application of $B_h$ and a negative $B_a$, the P25-15 petals soften and start to bend first due to the large heating power. During this time, the P5-15 petals are heated slowly and remain straight due to their lower temperature and high stiffness. With increasing heating time, the P5-15 petals start to soften and bend at 18 s and are eventually (at 32 s) fully actuated to lift the entire flower. After removing $B_h$ and cooling the flower down to room temperature, all petals are locked in their deformed shape. Fast transforming feature of M-SMPs is also demonstrated by switching the magnetic field direction during the actuation process (**Supplementary Video S3**). **Supplementary Video S4** shows a flower blooming-inspired sequential shape-transformation of an M-SMP system using P5-15, P15-15, P25-15.

**Sequential actuation for digital computing**

Soft active materials and structures have recently been explored for programmable mechanical computing due to its capability of integrating actuation and computing in soft bodies for potential applications in self-sensing of autonomous soft robots[39,40], nonlinear dynamics-enabled nonconventional computing[41], and mechanical logic circuits[42-44]. Taking M-SMPs' advantages of reversible actuation and shape locking, we demonstrate that M-SMPs can be used to design a sequential logic device, the D-latch, for storing one bit of information, which can be readily extended to a memory with arbitrary bits. The truth table for the D-latch logic is shown in **Fig. 3e**: when the input $E$ is 1, the output $Q$ keeps the same value as the input $D$; when the input $E$ is 0, the output $Q$ stays latched and is independent of the input $D$. We achieve this D-latch logic utilizing the controlled actuation of an M-SMP beam switch (**Fig. 3f & 3g**). The magnetic fields $B_h$ and $B_a$ work as inputs and the LED serves as the indicator of the output. The time-dependent actuation/locking of M-SMPs is interpreted as an RC delay circuit between $B_h$ and the D-latch, where the heating/cooling time of the M-SMP is regarded as the charging/discharging time of a capacitor (**Fig. S12**). When $B_h$ is on and the beam is unlocked ($T>T_g$, $E$=1), the downward $B_a$ ($D$=1) or upward $B_a$ ($D$=0) determines whether the circuit is closed or open, leading to the on ($Q$=1) or off ($Q$=0) state of the LED. When $B_h$ is off and the beam is locked ($T<T_g$, $E=0$), $B_a$ is no longer capable of actuating the beam and, consequently, cannot change the status of the LED. In other words, the previous state of $Q$ is stored in the system.

With the M-SMP-enabled D-latch system, we next design a sequential digital logic circuit as a three-bit memory, which contains three M-SMP beams (P5-15, P15-15, and P25-15) and three LEDs shown in **Fig. 3h** (**Methods, Fig. S13**). **Fig. 3i** shows the three-step logic for this three-bit memory, with $E_1$, $E_2$ and $E_3$ representing the input $E$ for P5-15, P15-15, and P25-15, respectively. During the operation, heating for 28 s unlocks all M-SMPs ($E_1$, $E_2$, $E_3$=1); heating for 12 s unlocks P5-15 and P25-15 ($E_1$=0, $E_2$, $E_3$=1); heating for 6 s only unlocks P25-15 ($E_1$, $E_2$=0, $E_3$=1). Followed by cooling and actuation (changing D), the M-SMP switches can lock their shapes and retain the output status. **Fig. 3j** shows the original state and output states for the three M-SMP switches indicated by the LEDs. In the first step, unlocking all M-SMPs ($E_1$, $E_2$, $E_3$=1) with D=1 changes the memory state from 0-0-0 to 1-1-1. After cooling and locking ($E_1$, $E_2$, $E_3$=0), we next unlock P15-15 and P25-15 ($E_1$=0, $E_2$, $E_3$=1) and switch D to 0, which changes the memory state from 1-1-1 to 1-0-0. In the third step after locking ($E_1$, $E_2$, $E_3$=0), we only unlock P25-15 ($E_1$, $E_2$=0, $E_3$=1) and switch D to 1 to finally change the memory state to 1-0-1 (**Supplementary Video S5**). This example demonstrates that by controlling the two inputs $B_h$ (E) and $B_a$ (D), we can erase and rewrite the information in the memory (**Table S3 & S4**). Theoretically, an electronic device with n-bit memory can be realized with $n$ M-SMPs with varying particle loadings. In this way, $2^n$ states can be achieved and stored with $n$ steps by manipulating two inputs. Additionally, we can tune the NdFeB particle loading and $T_g$ to provide more design flexibility for more complex computing systems using M-SMPs.

**Reprogrammable morphing radiofrequency antennas**

The ability to change the antenna shape on the fly provides the capability to either remotely deploy an antenna[45,46] or reconfigure its functionality[47-49]. Here, we demonstrate two morphing radio-frequency (RF) antennas that can rapidly, reversibly transform between on-demand shapes. The shape locking of M-SMPs allows the antennas to retain their actuated, functional shapes without the need for a constant application of external stimulation. **Fig. 4a** shows the design of a cantilever-based morphing monopole antenna (48 mm long). It can be reprogrammed to different magnetization profiles to transform into different shapes for deployable and reconfigurable antennas. Being magnetized along its longitudinal direction, gravity drives the cantilever to bend down (Down shape) upon heating. The Down shape can be actuated to the Up shape under $B_a$ = 20 mT (**Fig. 4b**). **Fig. 4c** shows the antenna works as a *deployable monopole antenna* due to its poor impedance ($S_{11}$ larger than the acceptable value, -10 dB[45,47]) in the Down shape but good $S_{11}$ value with a resonant frequency of 0.95 GHz in the Up shape. Moreover, this deployable antenna can be altered to a *reconfigurable antenna* by reprogramming the M-SMP's magnetization profile. Here, the same cantilever is remagnetized to have a sinusoidal shape with a height of 24 mm under $B_a$ = 80 mT (**Fig. 4b**, **Supplementary Video S6**, **Fig. S14**). **Fig. 4c** shows the resonant frequency of this antenna shifts from 0.95 GHz (Up shape) to 1.25 GHz (sinusoidal shape), representing a 32% change, with good agreement achieved between the simulation and experimental results. The radiation pattern simulations and polar plots are similar for all these configurations (**Fig. S14**), which is beneficial as a reconfigurable antenna.

Utilizing M-SMP's advantages of shape transformation and locking, the on-demand shape transformation from a planar state to a 3D structure can also be achieved. Here we design a tapered helical antenna to achieve frequency reconfigurability. The antenna is composed of a thin M-SMP substrate with printed conductive silver wire on its surface (**Fig. 4d**). The M-SMP substrate is magnetized in a stretched, spring-like configuration (**Fig. S15**) to realize the pop-up actuation with programmable heights and configurations under a controlled vertical $B_a$ (**Fig. 4e, Supplementary Video S7**). The simulation and experimental results in **Fig. 4f** show that the resonant frequency of the antenna can be readily tuned between 2.15 GHz and 3.26 GHz. The simulated radiation patterns at resonance with similar profiles shown in **Fig. 4g** indicate that the operating direction of the antenna remains constant, which is desirable for antenna applications. Due to the shape locking capability offered by the M-SMP, the reconfigured antenna can retain the new shape without assistance from the external field, which lowers the overall energy requirements. Using M-SMPs as a substrate material for a remotely controlled reconfigurable antenna is thus advantageous over the mechanically programmed antenna[33] and conventional magnetic-responsive antenna[50].

**Conclusion**

In summary, the reported magnetic shape memory polymer integrates reprogrammable, untethered, fast, and reversible shape transformation and shape locking into one system. Utilizing two types of embedded magnetic particles for inductive heating and actuation, the material can be effectively unlocked and locked for energy-efficient operations and functions as soft robotic grippers, sequential actuation devices, digital logic circuits, and deployable/reconfigurable antennas. With recent advances in simulation tools and 3D/4D printing for design optimization and fabrication of complex structures, these demonstrations suggest that the novel M-SMP can serve as a material platform for a wide range of applications, including biomedical devices for minimum invasive surgery, active metamaterials, morphological computing, autonomous soft robots, and reconfigurable, flexible electronics, etc.

**Methods**

**Preparation of the M-SMPs**

Our neat SMP is an acrylate-based amorphous polymer. The resin contains aliphatic urethane diacrylate (Ebecryl 8807, Allnex, GA), 2-Phenoxyethanol acrylate (Allnex, GA), isobornyl acrylate (Sigma-Aldrich, St. Louis, MO), and isodecyl acrylate (Sigma-Aldrich, St. Louis, MO) with a weight ratio of 0.7:60.2:30.1:9. A thermally-induced radical initiator (2,2'-Azobis(2-methylpropionitrile), 0.7 w%) is added for thermal curing. Additionally, 2 wt% of fumed silica with an average size of 0.2-0.3 μm (Sigma-Aldrich, St. Louis, MO) and 0.4 wt% of 2,2'-Azobis(2-methylpropionitrile) are added to ensure good mixing of the matrix resin with the magnetic particles. The composite is prepared by adding predetermined amounts of $Fe_3O_4$ (0-25 vol%) (particle size of 30 μm, Alpha Chemicals, MO, USA) and NdFeB magnetic particles (15 vol %) (average particle size of 25 μm, Magnequench, Singapore) in the matrix resin. The M-SMP composite is denoted as Px-y with x and y representing the $Fe_3O_4$ and NdFeB volume fractions, respectively. The mixture is manually mixed, degassed under vacuum, and then sandwiched between two glass slides with different spacings for thermal curing. The thermal curing is conducted by precuring at 80°C for 4 h and postcuring at 120°C for 0.5 h. The cured composite films are magnetized and remagnetized under impulse magnetic fields (~1.5 T for first magnetization and 5.5 T for remagnetization) generated by an in-house built impulse magnetizer.

**Electromagnetic coil system for actuation and inductive heating**

We use a pair of in-house built electromagnetic coils with coil spacing of 74 mm for actuating. The two coils are connected in series and powered by a custom programmable power supply with up to 15 A output current. The coils can generate a central magnetic field with a ratio of 7 mT/A. A water-cooled solenoid is connected to an LH-15A high-frequency induction heater to generate an AC magnetic field with a frequency of 60 kHz and a magnetic field ranging from 10 mT to 60 mT.

**Physical properties characterization**

Uniaxial tension tests are conducted on a dynamic mechanical analysis (DMA) tester (DMA 850, TA Instruments, New Castle, DE) at various temperatures. The thin film samples (dimension: about 20 mm × 3 mm × 0.6 mm) are stretched at a strain rate of 0.2/min. The dynamic thermomechanical properties are measured on the DMA tester. The strain is oscillated at a frequency of 1 Hz with a peak-to-peak amplitude of 0.1%. The temperature is ramped from 0°C to 120°C at the rate of 3°C/min. The shape memory tests are carried out on the DMA tester in the uniaxial tensile mode with controlled force. The thermal imaging video (Video S1) and temperature profiles (Fig. 2c & Fig. 3a) are recorded using a thermal imaging camera (Seek Thermal, Inc., Santa Barbara, CA). The dimension of M-SMPs used for the temperature profiles in Fig. 3a is 10 mm × 10 mm × 1 mm.

**Cantilever experiments**

The M-SMP film is cut into a strip with a length of 35 mm and width of 4.5 mm. Two acrylic pieces (15 mm × 7 mm × 2 mm) are used to clamp one end of the M-SMP strip to create a cantilever with a length of 20 mm.

**Gripper experiments**

Two M-SMP strips (47 mm × 5 mm × 0.5mm) are cut and glued together to form a cross shape. The dimension of each arm is 21 mm x 5 mm. The four-arm gripper is softened and mechanically deformed to a grasp shape (diameter: 15 mm). The gripper is then cooled down and magnetized along the direction shown in Fig. S9. After magnetization, a glass rod holder is glued to the central part of the gripper.

**Flower-like structure experiments**

The flower-like structure has two types of petals, one is P5-15 and one is P25-15. The dimensions of P5-15 and P25-15 petals are shown in Fig. S11. The petals are cut with the assist of acrylic molds. The individual petals are magnetized along the outward radial direction. The central part is 3D-printed using a commercial rigid resin using a Formlabs Form2 3D printer (Formlabs, Somerville, MA, USA). The petals are then glued to the central part.

**Sequential logic circuit experiments**

All beams used as switches in the sequential logic circuit have the dimension of 20 mm × 5 mm × 0.8mm. Each beam is fixed at one end to the printed circuit. Silver paste (Dupont ME603) is uniformly painted on the bottom surface of the beams and cured at 80°C for 20 min. The LED leads, the fixed end of beams, and the copper wires for connecting the power supply are all attached to the printed circuit using silver paste. Finally, the assembled circuit is cured at 80°C for another 20 min.

**Single cantilever-based antenna experiments**

The M-SMP film is cut into a strip with dimension of 50 mm × 10 mm × 0.25 mm. The designed silver band has a width of 6 mm and a length of 50 mm. Silver paste is painted on one side of the strip and cured at 80°C for 20 min. One end of the cantilever-based antenna sample is glued to a 3D-printed PLA base.

**Helical antenna experiments**

The helical antenna is fabricated with a 3D-printed PVA mold using an Ultimaker S5. The mold is filled with the M-SMP resin mixture and sandwiched between two glass slides for thermal curing. The curing reaction is conducted by precuring at 80°C for 4 h and post-treatment at 120°C for 0.5 h. The PVA mold is then dissolved using water. The cured sample is heated until soft, deformed to the shape as shown in Fig. S15, and magnetized along the height direction.

**Antenna simulation and measurements**

The antenna is transformed and locked to the expected actuated shape and is fed by a 50 Ω coaxial probe. The antenna's return loss ($S_{11}$) is measured using a Vector Network Analyzer (VNA). In all experiments, the antenna is connected to a 50 Ω SMA connector on a 300 mm x 300 mm aluminum ground plane. After achieving the desired antenna shape using $B_h$ and $B_a$, the feed pin of the SMA connector is connected to the conductive silver lines on the antenna, exciting the antenna for measurements. The bandwidths of interest during the measurement are from 0.5 GHz to 2 GHz for the cantilever-based antennas and 2 GHz to 4 GHz for the helical antenna. All antenna simulations are conducted using ANSYS Electromagnetic Suite V19.10 HFSS.

# References

1  Lendlein, A. & Kelch, S. Shape-memory polymers. *Angew Chem Int Edit* **41**, 2034-2057 (2002).
2  Jeon, S.-J., Hauser, A. W. & Hayward, R. C. Shape-Morphing Materials from Stimuli-Responsive Hydrogel Hybrids. *Accounts of Chemical Research* **50**, 161-169 (2017).
3  Liu, Y., Genzer, J. & Dickey, M. D. "2D or not 2D": Shape-programming polymer sheets. *Progress in Polymer Science* **52**, 79-106 (2016).
4  Jascha, U. S., Hortense Le, F., Paolo, E., André, R. S. & Andres, F. A. Programmable snapping composites with bio-inspired architecture. *Bioinspiration & Biomimetics* **12**, 026012 (2017).
5  Erb, R. M., Sander, J. S., Grisch, R. & Studart, A. R. Self-shaping composites with programmable bioinspired microstructures. *Nature communications* **4**, 1712 (2013).
6  Hu, W., Lum, G. Z., Mastrangeli, M. & Sitti, M. Small-scale soft-bodied robot with multimodal locomotion. *Nature* **554**, 81 (2018).
7  Huang, H.-W., Sakar, M. S., Petruska, A. J., Pané, S. & Nelson, B. J. Soft micromachines with programmable motility and morphology. *Nature Communications* **7**, 12263 (2016).
8  Yuan, J. *et al.* Shape memory nanocomposite fibers for untethered high-energy microengines. *Science* **365**, 155-158 (2019).
9  Zhang, Y.-F. *et al.* Fast-Response, Stiffness-Tunable Soft Actuator by Hybrid Multimaterial 3D Printing. *Advanced Functional Materials* **29**, 1806698 (2019).
10  Felton, S., Tolley, M., Demaine, E., Rus, D. & Wood, R. A method for building self-folding machines. *Science* **345**, 644-646 (2014).
11  Faber, J. A., Arrieta, A. F. & Studart, A. R. Bioinspired spring origami. *Science* **359**, 1386-1391 (2018).
12  Qin, J. *et al.* Injectable superparamagnetic ferrogels for controlled release of hydrophobic drugs. *Advanced Materials* **21**, 1354-1357 (2009).
13  Kobayashi, K., Yoon, C., Oh, S. H., Pagaduan, J. V. & Gracias, D. H. Biodegradable thermomagnetically responsive soft untethered grippers. *ACS applied materials & interfaces* **11**, 151-159 (2018).
14  Hosseinidoust, Z. *et al.* Bioengineered and biohybrid bacteria-based systems for drug delivery. *Advanced drug delivery reviews* **106**, 27-44 (2016).
15  Mostaghaci, B., Yasa, O., Zhuang, J. & Sitti, M. Bioadhesive bacterial microswimmers for targeted drug delivery in the urinary and gastrointestinal tracts. *Advanced Science* **4**, 1700058 (2017).
16  Ware, T. H., McConney, M. E., Wie, J. J., Tondiglia, V. P. & White, T. J. Voxelated liquid crystal elastomers. *Science* **347**, 982-984 (2015).
17  Donovan, B. R., Matavulj, V. M., Ahn, S.-k., Guin, T. & White, T. J. All-Optical Control of Shape. *Advanced Materials* **31**, 1805750 (2019).
18  Sydney Gladman, A., Matsumoto, E. A., Nuzzo, R. G., Mahadevan, L. & Lewis, J. A. Biomimetic 4D printing. *Nat Mater* **15**, 413-418 (2016).
19  Kim, Y., Yuk, H., Zhao, R., Chester, S. A. & Zhao, X. Printing ferromagnetic domains for untethered fast-transforming soft materials. *Nature* **558**, 274-279 (2018).
20  Xie, T. Tunable polymer multi-shape memory effect. *Nature* **464**, 267-270 (2010).
21  Hu, J. L., Zhu, Y., Huang, H. H. & Lu, J. Recent advances in shape-memory polymers: Structure, mechanism, functionality, modeling and applications. *Progress in Polymer Science* **37**, 1720-1763 (2012).

22   Meng, H. & Li, G. Q. A review of stimuli-responsive shape memory polymer composites. *Polymer* **54**, 2199-2221 (2013).

23   Kim, J. *et al.* Programming magnetic anisotropy in polymeric microactuators. *Nature Materials* **10**, 747 (2011).

24   Nguyen, V. Q., Ahmed, A. S. & Ramanujan, R. V. Morphing soft magnetic composites. *Advanced Materials* **24**, 4041-4054 (2012).

25   Liu, J. A.-C., Gillen, J. H., Mishra, S. R., Evans, B. A. & Tracy, J. B. Photothermally and magnetically controlled reconfiguration of polymer composites for soft robotics. *Science Advances* **5**, eaaw2897 (2019).

26   Jackson, J. A. *et al.* Field responsive mechanical metamaterials. *Science Advances* **4**, doi:ARTN eaau6419 10.1126/sciadv.aau6419 (2018).

27   Lum, G. Z. *et al.* Shape-programmable magnetic soft matter. *Proceedings of the National Academy of Sciences* **113**, E6007-E6015 (2016).

28   Zhao, R., Kim, Y., Chester, S. A., Sharma, P. & Zhao, X. Mechanics of hard-magnetic soft materials. *Journal of the Mechanics and Physics of Solids* **124**, 244-263 (2019).

29   Xu, T., Zhang, J., Salehizadeh, M., Onaizah, O. & Diller, E. Millimeter-scale flexible robots with programmable three-dimensional magnetization and motions. *Science Robotics* **4**, eaav4494 (2019).

30   Wallin, T., Pikul, J. & Shepherd, R. 3D printing of soft robotic systems. *Nature Reviews Materials* **3**, 84 (2018).

31   Shintake, J., Cacucciolo, V., Floreano, D. & Shea, H. Soft Robotic Grippers. *Advanced Materials* **30**, doi:ARTN 1707035 10.1002/adma.201707035 (2018).

32   Al‐Dahleh, R., Shafai, C. & Shafai, L. Frequency‐agile microstrip patch antenna using a reconfigurable mems ground plane. *Microwave and optical technology letters* **43**, 64-67 (2004).

33   Fu, H. *et al.* Morphable 3D mesostructures and microelectronic devices by multistable buckling mechanics. *Nature materials* **17**, 268 (2018).

34   Zhao, Q., Qi, H. J. & Xie, T. Recent progress in shape memory polymer: New behavior, enabling materials, and mechanistic understanding. *Progress in Polymer Science* **49-50**, 79-120 (2015).

35   Lendlein, A. & Gould, O. E. C. Reprogrammable recovery and actuation behaviour of shape-memory polymers. *Nature Reviews Materials* (2019).

36   Kumar, U. N., Kratz, K., Heuchel, M., Behl, M. & Lendlein, A. Shape‐Memory Nanocomposites with Magnetically Adjustable Apparent Switching Temperatures. *Advanced Materials* **23**, 4157-4162 (2011).

37   Wang, L. *et al.* Reprogrammable, magnetically controlled polymeric nanocomposite actuators. *Materials Horizons* **5**, 861-867 (2018).

38   Liu, Y., Shaw, B., Dickey, M. D. & Genzer, J. Sequential self-folding of polymer sheets. *Science Advances* **3**, e1602417 (2017).

39   Polygerinos, P. *et al.* Soft robotics: Review of fluid‐driven intrinsically soft devices; manufacturing, sensing, control, and applications in human‐robot interaction. *Advanced Engineering Materials* **19**, 1700016 (2017).

40   Treml, B. E. *et al.* Autonomous motility of polymer films. *Advanced Materials* **30**, 1705616 (2018).

41   Nakajima, K., Hauser, H., Li, T. & Pfeifer, R. Exploiting the dynamics of soft materials for machine learning. *Soft robotics* **5**, 339-347 (2018).

42  Devaraju, N. S. G. K. & Unger, M. A. Pressure driven digital logic in PDMS based microfluidic devices fabricated by multilayer soft lithography. *Lab on a Chip* **12**, 4809-4815 (2012).

43  Treml, B., Gillman, A., Buskohl, P. & Vaia, R. Origami mechanologic. *P Natl Acad Sci USA* **115**, 6916-6921 (2018).

44  Preston, D. J. *et al.* Digital logic for soft devices. *Proceedings of the National Academy of Sciences* **116**, 7750-7759 (2019).

45  Yao, S., Liu, X., Gibson, J. & Georgakopoulos, S. V. in *2015 IEEE International Symposium on Antennas and Propagation & USNC/URSI National Radio Science Meeting.*  2215-2216 (IEEE).

46  Costantine, J., Tawk, Y., Christodoulou, C. G., Banik, J. & Lane, S. CubeSat deployable antenna using bistable composite tape-springs. *IEEE Antennas and Wireless Propagation Letters* **11**, 285-288 (2012).

47  Kovitz, J. M., Rajagopalan, H. & Rahmat-Samii, Y. Design and implementation of broadband MEMS RHCP/LHCP reconfigurable arrays using rotated E-shaped patch elements. *IEEE Transactions on Antennas and Propagation* **63**, 2497-2507 (2015).

48  Oliveri, G., Werner, D. H. & Massa, A. Reconfigurable electromagnetics through metamaterials—A review. *Proceedings of the IEEE* **103**, 1034-1056 (2015).

49  Bernhard, J. T. Reconfigurable antennas. *Synthesis lectures on antennas* **2**, 1-66 (2007).

50  Langer, J.-C., Zou, J., Liu, C. & Bernhard, J. T. Micromachined reconfigurable out-of-plane microstrip patch antenna using plastic deformation magnetic actuation. *IEEE Microwave and Wireless Components Letters* **13**, 120-122 (2003).

**Acknowledgement**

R.Z. acknowledges the Haythornthwaite Foundation Research Initiation Grant and the OSU Emergent Materials Research Grant. H.J.Q acknowledges the support of an AFOSR grant (FA9550-19-1-0151; Dr. B.-L. "Les" Lee, Program Manager). F.Y.Y acknowledges the support from US Department of Energy under Grant No. DE-SC0001304. J.M.K. acknowledges the Georgia Tech Research Institute's independent research and development HIVE program under the direction of Mr. Benjamin Riley. The authors would like to thank Dr. Ken Allen and Dr. Todd Lee for their review of this article.

**Author Contributions**

R.Z. and H.J.Q. conceived the concept and designed the study. X.K. and Q.Z. developed the materials. Q.Z. and F.Y.Y. characterized the magnetic properties of the materials. X.K., S.W., and S.M.M. characterized the thermomechanical properties and shape memory performance of the materials. Q.Z. and S.W. designed the magnetic actuation experiments and Q.Z., S.W., and RD. Z. conducted these experiments. J.W., J.M.K., and Q.Z. designed morphing antennas. J.W. and J.M.K. conducted the antenna simulations and experiments. All the authors read the manuscript.

**Competing Financial Interests**

The authors declare no competing financial interests.

**Figures and Figure Captions**

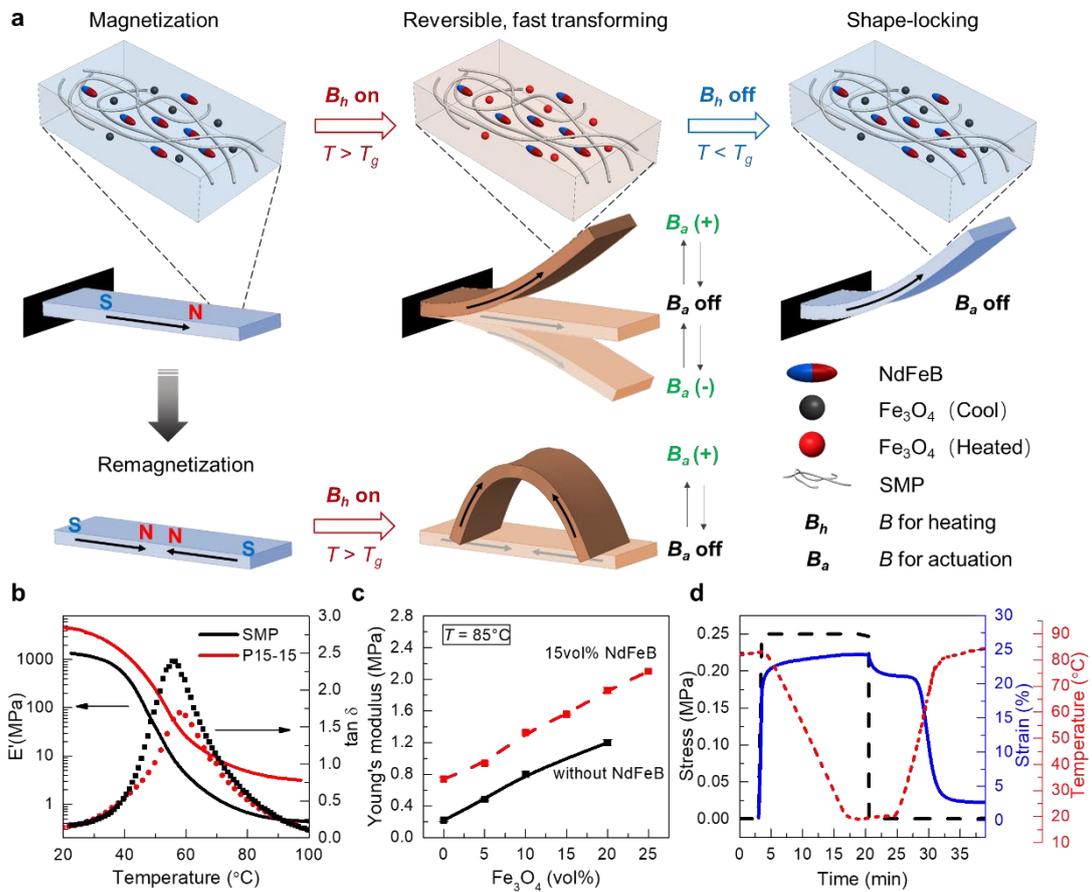

**Figure 1. Schematics and properties of magnetic shape memory polymers (M-SMPs).** (a) Working mechanism of M-SMPs. (b) Storage modulus and tan δ versus temperature for the neat SMP and P15-15 (M-SMP with 15 vol% $Fe_3O_4$ and 15 vol% NdFeB). (c) Effect of NdFeB and $Fe_3O_4$ particle loadings on the Young's modulus of the M-SMP at 85°C. (d) Shape memory performance of P15-15 (black dashed line: stress; blue solid line: strain; red dotted line: temperature).

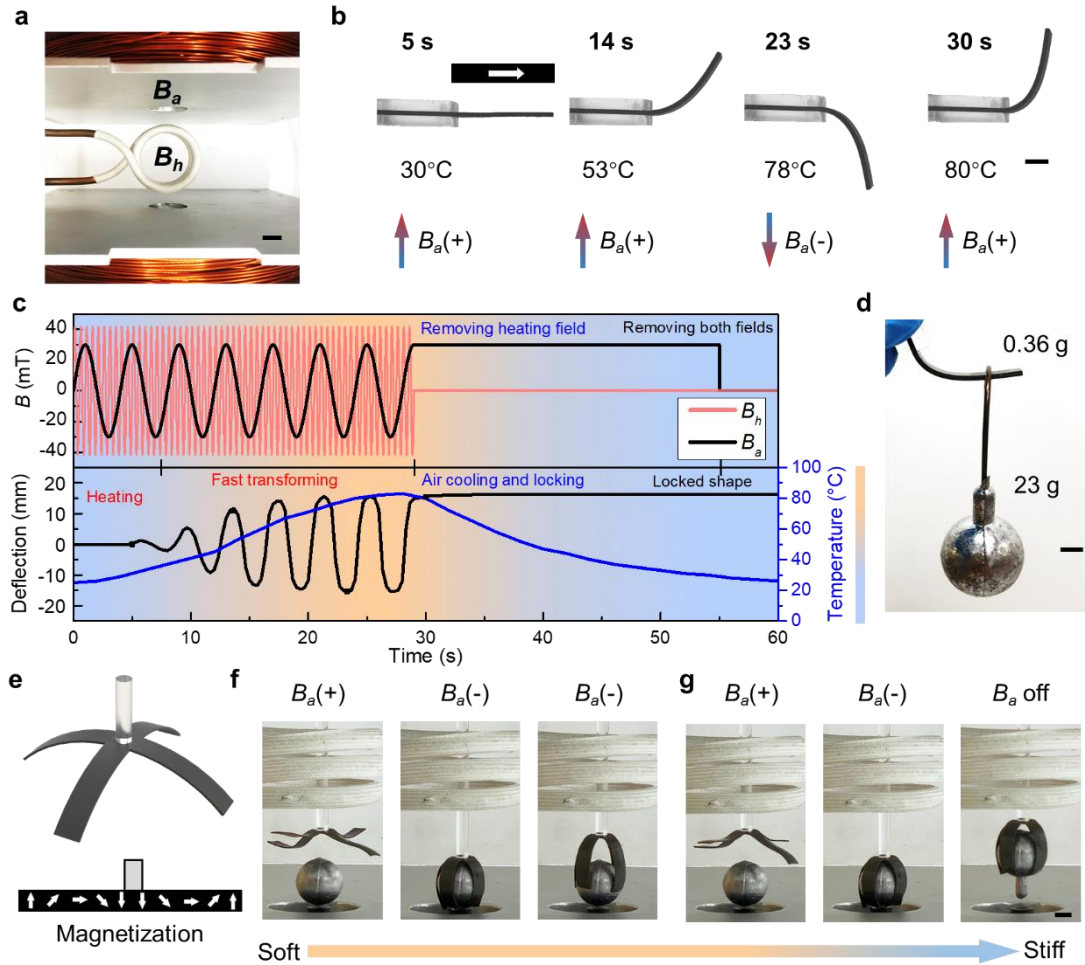

**Figure 2. Fast-transforming and shape locking of M-SMPs via superimposed magnetic fields.** (a) Experimental setup for the superimposed magnetic fields: the two parallel electric coils are used to generate the actuation magnetic field, $B_a$; the solenoid coil in the middle is used to generate the heating magnetic field, $B_h$. Scale bar: 15 mm. (b) Cantilever bending and shape locking. Scale bar: 5 mm. (c) Magnetic field profiles of $B_a$ and $B_h$ and beam deflection and temperature with respect to time. The gradient background color illustrates the time-dependent temperature change with the scale bar on the side. (d) Locked bending beam carrying a weight (23g) 64 times heavier than its own weight (0.36g). (e) Design and magnetization profile of a four-arm M-SMP gripper (0.47g). (f-g) M-SMP gripper lifting a lead ball (23g) without (f) and with (g) shape locking. Scale bar: 5 mm.

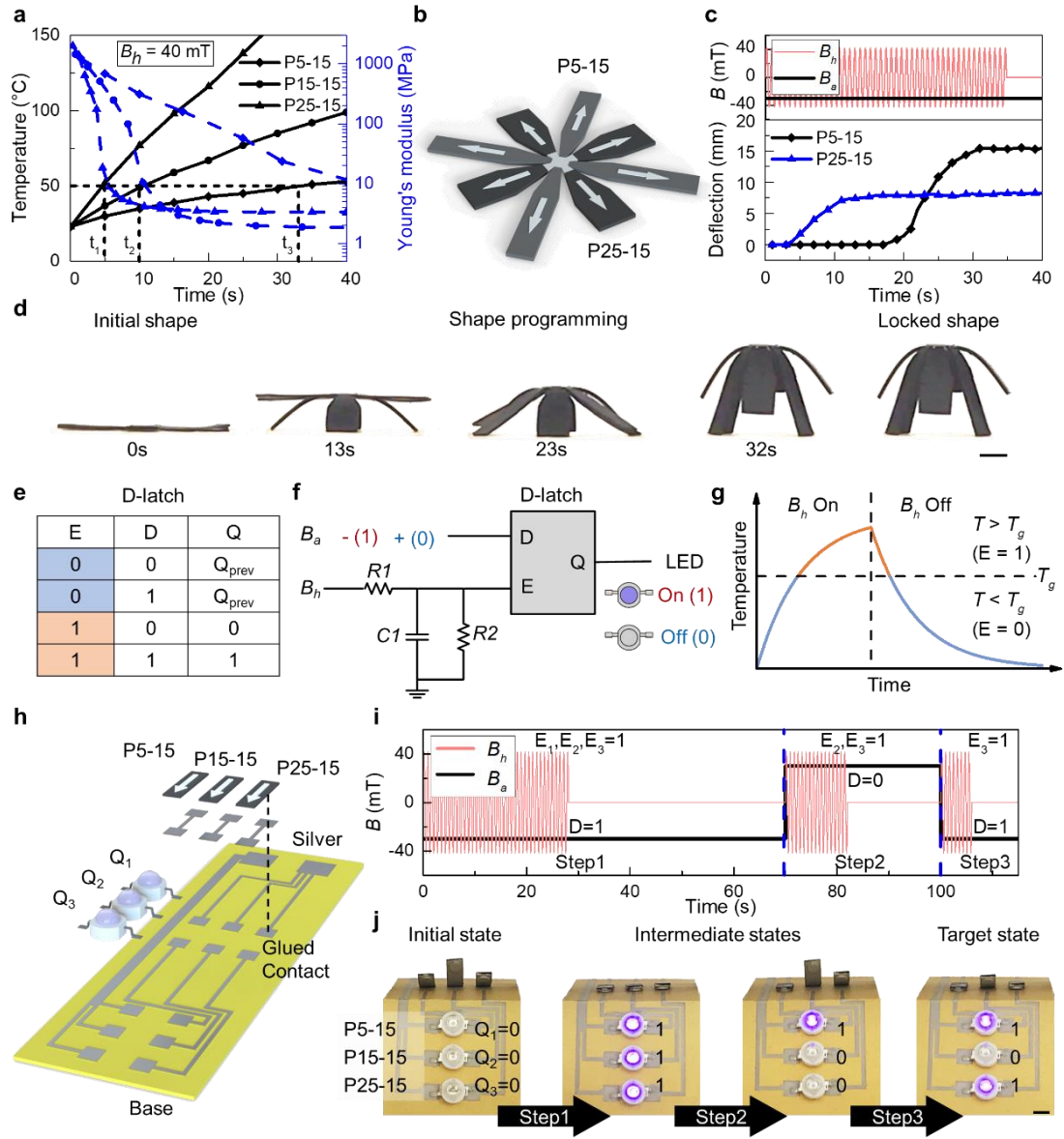

**Figure 3. Sequential actuation of M-SMPs and its application as digital logic circuits.** (a) Temperature and corresponding Young's moduli of three M-SMPs containing different $Fe_3O_4$ loadings. (b) Design of a flower-like structure using P5-15 and P25-15 M-SMPs. (c) Magnetic field profiles ($B_a$ and $B_h$) and deflection of the sequentially actuated M-SMPs with respect to time. (d) Sequential shape transforming and shape locking. (e) Truth table for a D-latch. (f) Schematic of an M-SMP D-latch logic with two magnetic fields ($B_a$ and $B_h$) serving as input and LED state as output. (g) Relationship between $B_h$ and the enabled input E of the D-latch. (h) Design of the sequential logic circuit using M-SMPs with different $Fe_3O_4$ loadings (P5-15, P15-15, and P25-15). (i) Magnetic control for a sequential logic circuit with three steps and tunable outputs. (j) LED indications for four different output states. Scale bars in (d) and (j): 5 mm.

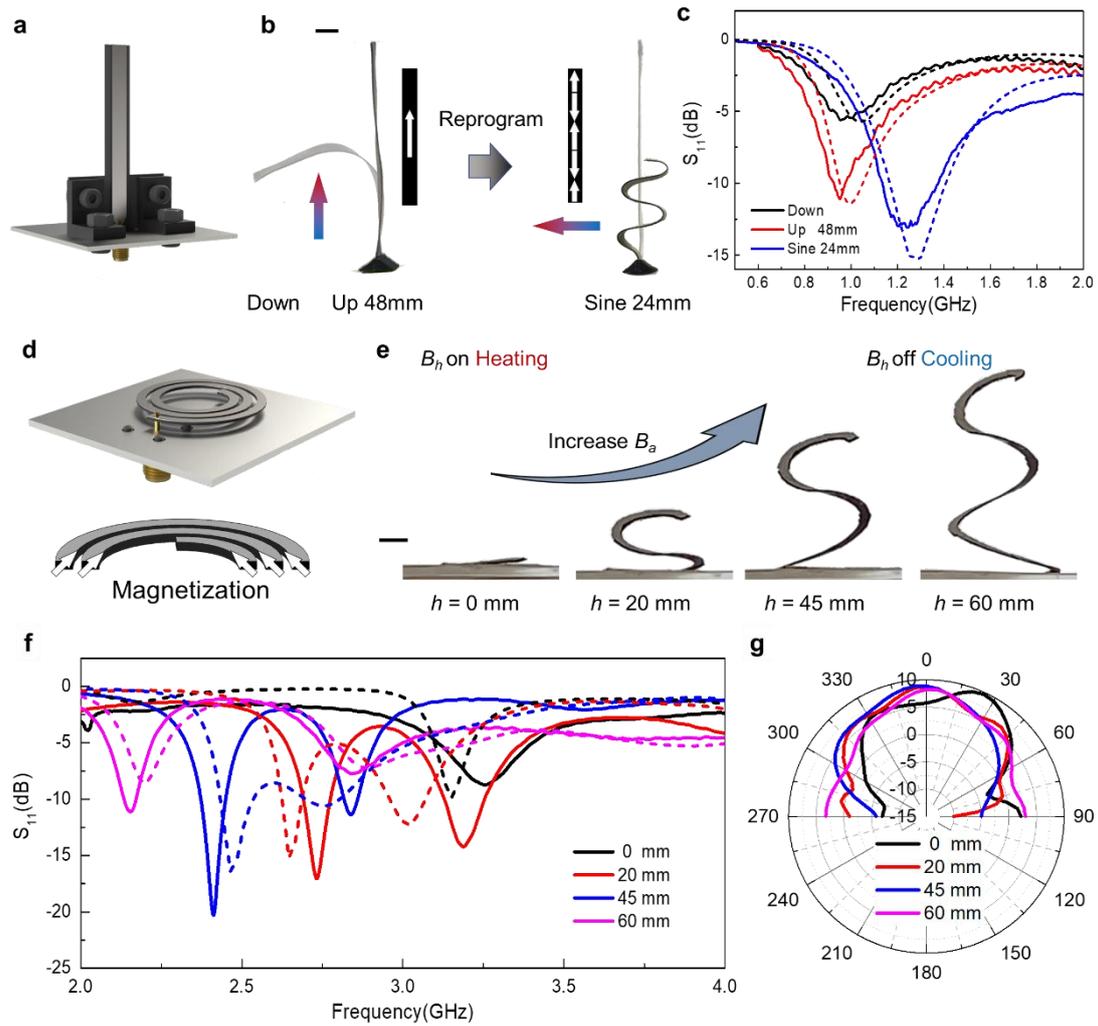

**Figure 4. Application of M-SMPs for morphing antennas.** (a) Schematic of a single-cantilever monopole antenna. (b) Cantilever antenna with two different magnetization profiles by reprogramming. (c) Experimental (solid lines) and simulation (dashed lines) results of the $S_{11}$ spectrum. (d) Schematic and magnetization profile of a reconfigurable helical antenna. (e) Actuation of the helical antenna under different $B_a$. (f) Experimental (solid lines) and simulation (dashed lines) results of $S_{11}$ band for the reconfigurable helical antenna at different heights. (g) 2D polar plot of the simulated radiation patterns of the helical antenna at different heights. Scale bars in (b) and (e): 5 mm.